\documentclass[pre,showpacs,floatfix,twocolumn]{revtex4-1}
\usepackage{amsfonts}
\usepackage{amsmath}
\usepackage{amssymb}
\usepackage{graphicx}
\usepackage{epsfig}

\begin{document}

\title{Time-dependent Monte Carlo simulations of the critical and Lifshitz
points of the ANNNI model.}
\author{Roberto da Silva}
\email{rdasilva@if.ufgrs.br}
\affiliation{Instituto de Fisica, Universidade Federal do Rio Grande do Sul,
Av. Bento Gon\c{c}alves, 9500 - CEP 91501-970, Porto Alegre, Rio Grande do
Sul, Brazil}
\author{Nelson Alves Jr.}
\email{nalves@if.usp.br}
\affiliation{ Faculdade de Filosofia, Ci\^{e}ncias e Letras de Ribeir\~{a}o
Preto, Universidade de S\~{a}o Paulo, Avenida Bandeirantes, 3900 - CEP
14040-901, Ribeir\~{a}o Preto, S\~{a}o Paulo, Brazil}
\author{Jose Roberto Drugowich de Felicio}
\email{drugo@usp.br}
\affiliation{ Faculdade de Filosofia, Ci\^{e}ncias e Letras de Ribeir\~{a}o
Preto, Universidade de S\~{a}o Paulo, Avenida Bandeirantes, 3900 - CEP
14040-901, Ribeir\~{a}o Preto, S\~{a}o Paulo, Brazil}

\begin{abstract}
In this work, we study the critical behavior of second order points and
specifically of the Lifshitz point (LP) of a three-dimensional Ising model
with axial competing interactions (ANNNI model), using time-dependent Monte
Carlo simulations. First of all, we used a recently developed technique that
helps us localize the critical temperature corresponding to the best power
law for magnetization decay over time: $\left\langle M\right\rangle
_{m_{0}=1}\sim t^{-\beta /\nu z}\ $ which is expected of simulations
starting from initially ordered states. Secondly, we obtain original results
for the dynamic critical exponent $z$, evaluated from the behavior of the
ratio $F_{2}(t)=\left\langle M^{2}\right\rangle _{m_{0}=0}/\left\langle
M\right\rangle _{m_{0}=1}^{2}\sim t^{3/z}$, along the critical line up to
the LP. Finally, we explore all the critical exponents of the LP in detail,
including the dynamic critical exponent $\theta $ that characterizes the
initial slip of magnetization and the global persistence exponent $\theta
_{g}$ associated to the probability $P(t)$ that magnetization keeps its
signal up to time $t$. Our estimates for the dynamic critical exponents at
the Lifshitz point are $z=2.34(2)$ and $\theta _{g}=0.336(4)$, values very
different from the 3D Ising model (ANNNI model without the
next-nearest-neighbor interactions at $z$-axis, i.e., $J_{2}=0$) $z\approx
2.07$ and $\theta _{g}\approx 0.38$. We also present estimates for the
static critical exponents $\beta $ and $\nu $, obtained from extended
time-dependent scaling relations. Our results for static exponents are in
good agreement with previous works
\end{abstract}

\maketitle


\section{Introduction}

\subsection{Modulated systems and the Lifshitz point}

In condensed matter physics there are several models presenting spatial
modulated structures of some local property, such as, for instance, the
position of the particles, the magnetization and the charge density \cite%
{Bak1982}. Such modulation can be commensurate or incommensurate in relation
to the underlying lattice. A phase is named commensurate if the ratio
between the period of the modulation\ and the period of the lattice is a
rational number. Otherwise, the phase is named incommensurate. The basic
mechanism leading to modulation is the competition between interactions
favoring distinct orderings \cite{Toulouse1977}. For example, the modulated
structures observed in rare-earth metals \cite{Cooper1968} were interpreted
as a consequence of the competition generated by the spatially oscillatory
interaction RKKY \cite{Ruderman1954,Kasuya1956,Yosida1957}. In order to
explain the spatial modulation found in Erbium, Elliott introduced an Ising
model \cite{Elliot1961}, twenty years later named ANNNI
(Axial-Next-Nearest-Neighbor Ising) model \cite{Fisher1980}. It is one of
the simplest models able to exhibit a rich phase diagram containing a
complex region of spatially modulated phases \cite%
{Selke1988,Yeomans1988,Selke1992}. The model is defined by the Hamiltonian

\begin{equation*}
\begin{array}{lll}
\mathcal{H} & = & -\sum_{x,\ y,\ z}\left[ J_{0}(\sigma _{x+1,y,z}+\sigma
_{x,y+1,z})+J_{1}\sigma _{x,y,z+1}\right. \\ 
&  &  \\ 
&  & \left. +\ J_{2}\sigma _{x,y,z+2}\right] \ \sigma _{x,y,z}%
\end{array}%
\end{equation*}%
where $\sigma _{x,y,z}=\pm 1$ is an Ising spin variable at the site $(x,y,z)$
of a simple cubic lattice, $J_{0}$ is the nearest neighbor interaction in
the $xy$ planes, $J_{1}$and $J_{2}$ are the nearest- and
next-nearest-neighbor interactions in the $z$ direction.\textbf{\ }Here $%
J_{1}$\ and $J_{2}$\ compete and may have opposite or same signs. However,
when $J_{1}$\ and $J_{2}$\ have opposite signs, the competition between them
may give rise to the modulated phases.

\begin{figure}[h]
\begin{center}
\includegraphics[width=\columnwidth]{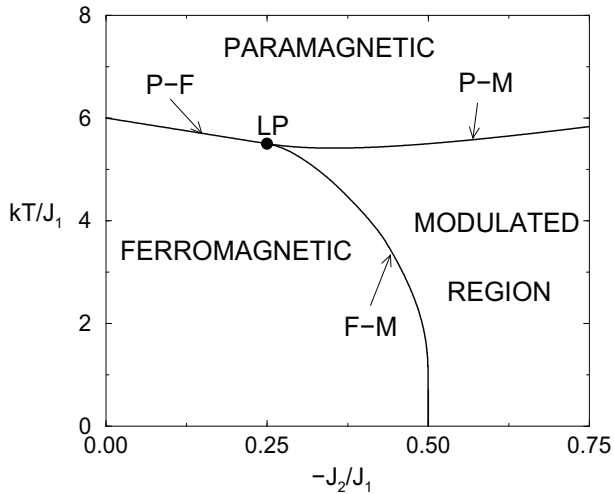}
\end{center}
\caption{Mean Field phase diagram of the ANNNI Model, displaying the
ferromagnetic phase, the paramagnetic phase and the modulated region. The
critical lines P-F and P-M meet the first order phase transition line F-M at
the Lifshitz point (LP) (see text). }
\label{phase_diagram}
\end{figure}

The mean-field phase diagram of the model displaying the main commensurate
phases in the plane of the reduced temperature $kT/J_{1}$ against the
competition parameter $-J_{2}/J_{1}$, shown in Fig. \ref{phase_diagram}, was
obtained in the beginning of the eighties \cite{Bak1980}\ \ and is divided
in three major regions: the modulated region (M), the paramagnetic (P) and
the ferromagnetic (F) phases. In this diagram, as well as in the whole
paper, we make $J_{0}=J_{1}$. The paramagnetic phase is separated from the
ferromagnetic phase and from the modulated region by critical lines P-F and
P-M respectively. On the other hand, \ a first-order transition (F-M) takes
place between the ferromagnetic and the modulated region. The critical line
P-M belongs to the universality class of the XY model, whereas the critical
line P-F presents Ising-like critical behavior \cite{Garel1976}. The
transition lines P-F, P-M and F-M meet at the Lifshitz point (LP),
introduced theoretically by Hornreich et al. \cite{Horneich1975}. The
location of the LP in the phase diagram of the ANNNI model was obtained for
the first time from high temperature series technique \cite{Redner1977}.
However, the precise location of the LP is difficult to obtain because,
close to it, one meets the challenging task of dealing with modulated phases
of very large periods. Thus, latter attempts to locate the LP were made.
Single spin-flip Monte Carlo simulations were performed along the P-F
critical line towards the LP \cite{Selke1978,Kaski1985}, thus avoiding the
modulated region. Despite the successful results obtained from these works,
it was only recently that the Lifshitz point was located with high precision 
\cite{Henkel2002}\ by means of a new variation of the cluster Wolf algorithm 
\cite{Wolf1989}. Concerning the critical properties of the LP, it was shown
that this multicritical point belongs neither to the universality class of
the XY model nor to the one of Ising model \cite{Horneich1975}. This fact
gave rise to much interest in the study of the particular critical behavior
of a LP. Thus, the critical properties of the LP found in the phase diagram
of the ANNNI model were widely studied by several approaches \cite%
{Hornreich1980,Mo1991}, among them $\epsilon -$expansions\cite{Horneich1975}%
, high temperature series technique \cite{Mo1991} and Monte Carlo
simulations \cite{Kaski1985}. Experimentally, the most complete results
concerning the LP were obtained for the magnetic compound MnP, which
exhibits a LP belonging to the universality class of the LP present in the
ANNNI model \cite{Beerra1980,Shapira1981,Bindilatti1989}. Here it is
important to mention that although this model has been widely explored by
methods from equilibrium statistical mechanics, there are no results for the
3D ANNNI model obtained by an approach that deals with MC simulations
performed far from equilibrium \cite{Li1996} that will be employed in this
paper. However we remark that several works in the literature have used
previously a dynamical approach to study the Lifshitz points. For example
the critical dynamics of relaxation of the model near the Lifshitz point has
been studied by the $\varepsilon -$expansion and the exponent $z$\ was
numerically estimated for uniaxial and biaxial cases \cite{Folk1978}.
Alternatively, in \cite{Basu2004} the growth of a order parameter is studied
when a system at Lifshitz point is quenched from the homogeneous disordered
state to the ordered state where correlation and structure factors after
quench in this system are analyzed via Renormalization Group method.
Numerical estimates of exponent $z$\ are supplied in this same reference.
Studies about dynamics of ANNNI model were developed but in two dimensional
versions of the model but such works do not explore critical properties. For
example, in \cite{Ala-Nissila1988}, \cite{Cheon2001}, the kinetics of
domain growth of the one and two-dimensional anisotropic ANNNI model was
explored via Monte Carlo methods using Glauber and Heat-Bath multispin
dynamics respectively.

In the next subsection we briefly present the non-equilibrium approach to be
used in our analysis.

\subsection{Non-equilibrium Critical dynamics}

The study of critical properties originated from statistical fluctuations of
spin systems became possible in non-equilibrium stage after the seminal
ideas of Janssen, Schaub and Schmittmann~\cite{Janssen1989} and Huse~\cite%
{Huse1989}. By quenching systems from high temperatures to the critical one,
they have shown that universality and scaling behavior appear even in the
early stages of time evolving, via renormalization group techniques and
numerical calculations respectively. Hence, by using short time dynamics,
one can often circumvent the well-known problem of the critical slowing down
that plagues investigations of the long-time regime.

Here, we briefly review finite size scaling in the dynamics relaxation of
spin systems. We present our alternative derivation of some power laws in
the short time dynamics context. Readers, who want a more complete review
about this topic, may want to read \cite{Zheng1998,Albano2011}.

This topic is based on time dependent simulations, and it constitutes an
important issue in the context of phase transitions and critical phenomena.
Such methods can be applied not only to estimate the critical parameters in
spin systems, but also to calculate the critical exponents (static and
dynamic ones) through different scaling relations by setting different
initial conditions.

The dynamic scaling relation obtained by Janssen \textit{et al.} for the 
\textit{k}-th moment of the magnetization, is written as 
\begin{equation}
\langle M^{k}\rangle (t,\tau ,L,m_{0})=b^{-k\beta /\nu }\langle M^{k}\rangle
(b^{-z}t,b^{1/\nu }\tau ,b^{-1}L,b^{x_{0}}m_{0})\text{,}
\label{mainshorttime}
\end{equation}%
where the arguments are: the time $t$; the reduced temperature $\tau
=(T-T_{c})/T_{c}$, with $T_{c}$ being the critical one, the lattice linear
size $L$ and initial magnetization $m_{0}$. Here, the operator $\langle
\ldots \rangle $ denotes averages over different configurations due to
different possible time evolutions from each initial configuration
compatible with a given $m_{0}$.

On the right-hand side of the equation, one has: an arbitrary spatial
rescaling factor $b$; an anomalous dimension $x_{0}$ related to $m_{0}$. The
exponents $\beta $ and $\nu $ are the equilibrium critical exponents
associated with the order parameter and the correlation length,
respectively. The exponent $z$ is the dynamic one, which characterizes the
time correlations in equilibrium. After choosing $b^{-1}L=1$, $T=$ $T_{c}$($%
\tau =0$), and $k=1$ we obtain $\langle M\rangle (t,L,m_{0})=L^{-\beta /\nu
}\langle M\rangle (L^{-z}t,L^{x_{0}}m_{0})$.

Denoting $u=tL^{-z}$ and $w=L^{x_{0}}m_{0}$, one has: $\langle M\rangle
(u,w)=\langle M\rangle (L^{-z}t,L^{x_{0}}m_{0})$. The derivative with
respect to $L$ is: 
\begin{eqnarray*}
\frac{\partial \langle M\rangle }{\partial L} &=&(-\beta /\nu )L^{-\beta
/\nu -1}\langle M\rangle (u,w)+ \\
&&L^{-\beta /\nu }\left[\frac{\partial \langle M\rangle }{\partial u}\frac{%
\partial u}{\partial L}+\frac{\partial \langle M\rangle }{\partial w}\frac{%
\partial w}{\partial L}\right],
\end{eqnarray*}%
where we have explicitly: $\partial u/\partial L=-ztL^{-z-1}$ and $\partial
w/\partial L=x_{0}m_{0}L^{x_{0}-1}$. In the limit $L\rightarrow \infty $, $%
\partial _{L}\langle M\rangle \rightarrow 0$, one has: $x_{0}w\frac{\partial
\langle M\rangle }{\partial w}-zu\frac{\partial \langle M\rangle }{\partial u%
}-\beta /\nu \langle M\rangle =0$. The separability of the variables $u$ and 
$w$ in $\langle M\rangle (u,w)=M_{1}(u)M_{2}(w)$ leads to $%
x_{0}wM_{2}^{\prime }/M_{2}=\beta /\nu +zuM_{1}^{\prime }/M_{2}$, where the
prime means the derivative with respect to the argument. Since the
left-hand-side of this equation depends only on $w$ and the right-hand-side
depends only on $u $, they must be equal to a constant $c$. Thus, $%
M_{1}(u)=u^{(c/z)-\beta /(\nu z)}$ and $M_{2}(w)=w^{c/x_{0}}$, resulting in $%
\left\langle M\right\rangle (u,w)=m_{0}^{c/x_{0}}L^{\beta /\nu }t^{(c-\beta
/\nu )/z}$. Returning to the original variables, one has: $\langle M\rangle
(t,L,m_{0})=m_{0}^{c/x_{0}}t^{(c-\beta /\nu )/z}$.

On one hand, choosing $c=x_{0}$ and calculating $\theta =(x_{0}-\beta /\nu
)/z$, at criticality ($\tau =0$), we obtain 
\begin{equation}
\langle M\rangle _{m_{0}}\sim m_{0}t^{\theta }  \label{initial slip}
\end{equation}%
corresponding to a regime under small initial magnetization. This can be
observed by a finite time scaling $b=t^{1/z}$ in equation (\ref%
{mainshorttime}), at critical temperature ($\tau =0$) which leads to $%
\left\langle M\right\rangle (t,m_{0})=t^{-\beta /(\nu z)}\langle M\rangle
(1,t^{x_{0}/z}m_{0})$. Defining $x=t^{x_{0}/z}m_{0}$, an expansion of the
averaged magnetization around $x=0$ results in: $\langle M\rangle
(1,x)=\langle M\rangle (1,0)+\left. \partial _{x}\langle M\rangle
\right\vert _{x=0}x+\mathcal{O}(x^{2})$. By construction $\langle M\rangle
(1,0)=0$, since $x=t^{x_{0}/z}m_{0}\ll 1$ and $\left. \partial _{x}\langle
M\rangle \right\vert _{x=0}$ is a constant, discarding the quadratic terms
we obtain the expected power law behavior $\langle M\rangle _{m_{0}}\sim
m_{0}t^{\theta }$. This anomalous behavior of initial magnetization is valid
only for a characteristic time scale $t_{\max }$ $\sim m_{0}^{-z/x_{0}}$.

On the other hand, the choice $c=0$ corresponds to the case which the system
does not depend on the initial trace and $m_{0}=1$ leads to simple power
law: 
\begin{equation}
\langle M\rangle _{m_{0}=1}\sim t^{-\beta /(\nu z)}  \label{decay_ferro}
\end{equation}%
that similarly corresponds to decay of magnetization for $t>t_{\max }$. The
evaluation of critical exponents $\theta $ and $\beta /(\nu z)$ via Monte
Carlo simulations concerns taking averages over different runs. In the
second case, simpler simulations are considered because since the system
starts from the ferromagnetic (ordered) initial state. However, the first
one is somewhat difficult to deal with, once it demands working with
prepared initial states with a precise value of $m_{0}$ (sharp preparation),
besides the delicate limit $m_{0}\rightarrow 0$.

An alternative way to determine this exponent was proposed by Tom\'{e} and
Oliveira \cite{Tome1998}, where it was shown that the time correlation
function of the order parameter also follows a power law scale form at the
short-time regime, i. e.

\begin{equation}
Q(t)=\left\langle M(0)M(t)\right\rangle \sim t^{\theta }.
\label{correlation}
\end{equation}

The main advantage in the use of Eq. (\ref{correlation}) is that one does
not need to fix precisely the initial order parameter $m_{0}$. The only
requirement is that $\left\langle m_{0}\right\rangle =0$, where $%
\left\langle (...)\right\rangle $ stands for the average of the quantity $%
\left( ...\right) $ over different initial configurations. Now, let us look
at the second moment of magnetization. Since the spin-spin correlation $%
\left\langle \sigma _{i}\sigma _{j}\right\rangle $ is negligible for $m_{0}=0
$, we have that for a fixed $t$, 
\begin{equation*}
\left\langle M^{2}\right\rangle _{m_{0}=0}=\frac{1}{L^{2d}}%
\sum\limits_{i=1}^{L^{d}}\left\langle \sigma _{i}^{2}\right\rangle +\frac{1}{%
L^{2d}}\sum\limits_{i}^{L^{d}}\left\langle \sigma _{i}\sigma
_{j}\right\rangle \approx L^{-d}
\end{equation*}%
and by considering the scaling relation (with $b=t^{1/z}$) for the second
moment of magnetization, we have according to \ref{mainshorttime}:

\begin{equation}
\begin{array}{lll}
\left\langle M^{2}\right\rangle _{m_{0}=0}(t,L) & \approx & t^{\frac{-2\beta 
}{\nu z}}\left\langle M^{2}\right\rangle _{m_{0}=0}(1,bL) \\ 
\  & \  & \  \\ 
\  & = & t^{\frac{-2\beta }{\nu z}}(bL)^{-d} \\ 
\  & \  & \  \\ 
\  & \sim & t^{(d-\frac{2\beta }{\nu })/z}%
\end{array}
\label{M2}
\end{equation}%
where $d$ is the system dimension.

Using Monte Carlo simulations, many authors have obtained the dynamic
exponents $\theta $ and $z$ as well as the static ones $\beta $ and $\nu $,
and other specific exponents for several models: Baxter-Wu~\cite%
{Arashiro2003}, 2, 3 and 4-state Potts~\cite{SilvaI2002,daSilva2004}, Ising
with multispin interactions~\cite{Simoes2001}, Ising with competing
interactions \cite{AlvesJr2003}, models with no defined Hamiltonian (celular
automata)~\cite{daSilva2005}, models with tricritical point~\cite%
{SilvaII2002}, Heisenberg~\cite{Fernandes2006}, protein folding~\cite%
{Arashiro3} and so on.

The sequence to determine the static exponents from short time dynamics is:
first we determine $z$, performing Monte Carlo simulations that mixes
initial conditions~\cite{SilvaI2002}, and consider the power law for the
cumulant 
\begin{equation}
F_{2}(t)=\frac{\left\langle M^{2}\right\rangle _{m_{0}=0}}{\left\langle
M\right\rangle _{m_{0}=1}^{2}}\sim t^{d/z}\;.  \label{z}
\end{equation}

This ratio has proven to be useful for the calculation of the exponent $z$
for the several spin models governed by Boltzmann-Gibbs Statistical
Mechanics but its application also includes models with spin-flip based on
generalized statistics \cite{daSilva2012}. In this technique, graphs of $\ln
F_{2}$ against $\ln t$ lay on the same straight line for different lattice
sizes, without any re-scaling in time, yielding more precise estimates for $z
$. Although it seems clear from Eq. (\ref{z}), it is worth to stress here
that two independent runs are necessary in order to calculate the ratio $%
F_{2}$: In one of them $m_{0}=0$, while in the other one $m_{0}=1$.

Equations (\ref{correlation}),(\ref{initial slip}), and (\ref{z}) solve the
problem in determining the dynamic critical exponents $\theta $ and $z$. But
the ability of short-time Monte Carlo simulations goes beyond the evaluation
of dynamic critical exponents, in the sense that this technique also allows
us to obtain the static critical exponents which will be discussed in
sequence.

Starting from $m_{0}=1$, we have the expected power law described by Eq.(\ref%
{decay_ferro}), so considering $\ln M(t,\tau )$, we must expect 
\begin{equation}
\left. \frac{\partial \ln M(t,\tau )}{\partial \tau }\right\vert _{\tau
=0}\sim t^{\frac{1}{\nu z}}  \label{derivative}
\end{equation}%
which is obtained by differentiating the quantity $\ln M(t,\tau )$ in
relation to reduced temperature $\tau =(T-T_{c})/T_{c}$. When dealing with
Monte Carlo simulations, the partial derivative is approximated in first
order by the difference 
\begin{equation}
\left. \frac{\partial \ln M(t,\tau )}{\partial \tau }\right\vert _{\tau
=0}\approx \frac{1}{2\varepsilon }\ln \left[ \frac{M(t,T_{c}+\varepsilon )}{%
M(t,T_{c}-\varepsilon )}\right]  \label{aproximation}
\end{equation}%
where $\varepsilon <<1$. It is clear from Eq. (\ref{aproximation}) above
that two independent simulations are necessary to obtain the exponent $1/\nu
z$: one of them evolves at the temperature $T_{c}+\varepsilon $, whereas the
other one evolves at $T_{c}-\varepsilon $.

Therefore, with $\widehat{z}$ estimated from Eq.(\ref{z}) we obtain the
estimated $\widehat{\nu }$ according to $\widehat{(\nu z)^{-1}}$ obtained
from Eq.(\ref{derivative}), by calculating the product:%
\begin{equation*}
\widehat{\nu }=\left[ \widehat{z}\cdot \widehat{(\nu z)^{-1}}\right] ^{-1}
\end{equation*}%
where the uncertainty in $\widehat{\nu }$ is calculated through uncertainty
of $z$ ($\sigma _{z}$) and the uncertainty of $\widehat{(\nu z)^{-1}}$ ($%
\sigma _{(\nu z)^{-1}}$), previously obtained by statistics over $n_{s}$
different seeds, according to the error propagation equation 
\begin{equation*}
\sigma _{\nu }=\sqrt{\frac{\sigma _{z}^{2}}{\widehat{z}^{4}\widehat{(\nu
z)^{-1}}^{2}}+\frac{\sigma _{(\nu z)^{-1}}^{2}}{\widehat{z}^{2}\widehat{(\nu
z)^{-1}}^{4}}}
\end{equation*}

Finally, in order to obtain an estimate $\widehat{\beta }$, we firstly
estimate $\widehat{(\beta /\nu z)}$ obtained from the magnetization decay (%
\ref{decay_ferro}) and after we perform the product 
\begin{equation*}
\widehat{\beta }=\widehat{(\beta /\nu z)}\cdot \left[ \widehat{(\nu z)^{-1}}%
\right] ^{-1}
\end{equation*}%
and the propagated error is directly calculated as a function of the
respective uncertainties:%
\begin{equation*}
\sigma _{\beta }=\sqrt{\frac{\sigma _{(\beta /\nu z)}^{2}}{\widehat{(\nu
z)^{-1}}^{2}}+\frac{\widehat{(\beta /\nu z)}^{2}\sigma _{(\nu z)^{-1}}^{2}}{%
\widehat{(\nu z)^{-1}}^{4}}}
\end{equation*}

Here, it is important to mention that the ratio $\beta /\nu $ already should
have been previously calculated by the evaluated $\widehat{(\beta /\nu z)}$
from Eq.(\ref{decay_ferro}) and $\widehat{z}$ from Eq.(\ref{z}) such that:%
\begin{equation}
\widehat{\beta /\nu }=\widehat{(\beta /\nu z)}\cdot \widehat{z}  \label{eta}
\end{equation}

Not only quantities related to the moments of the magnetization can explain
the non-equilibrium aspects of phase transitions and critical phenomena, but
also those related to the first passage time probabilities and variations of
this topic (see e.g \cite{Mglobal96},\cite{Majumdar96}). By considering this
approach and under the same nonequilibrium conditions, a new exponent was
initially proposed by \cite{Mglobal96}: the global persistence exponent $%
\theta _{g}$. It is related to the probability $P(t)$ that the global order
parameter has not changed sign up to time $t$ after a quench to $T_{c}$,
according to expected power law behavior%
\begin{equation}
P(t)\sim t^{-\theta _{g}}\text{.}  \label{global_persistence}
\end{equation}%
As argued by Majumdar \textit{et al.} \cite{Mglobal96}, if the dynamics of
the global order parameter is described by a Markovian process, $\theta _{g}$
is not a new independent exponent and it can be related to other critical
exponents, 
\begin{equation}
\theta _{g}z=\omega z-d+1-\eta /2\,.  \label{global}
\end{equation}%
where $\omega $ is the autocorrelation exponent in the expected power law 
\cite{Janssen1992}: $A(t,t%
{\acute{}}%
=0)_{m_{0}=0}=(1/L^{d})\left\langle \sum\nolimits_{i}\sigma _{i}(t)\sigma
_{i}(0)\right\rangle \sim t^{-\omega }$, where $\sigma _{i}(t)$ is the value
of the spin variable $s_{i}$ at the site $i$ of a d-dimensional system of
linear size L, assumed, at instant $t$, $z$ is the dynamic critical exponent
defined as $\tau \sim \xi ^{z}$ , where $\tau $ and $\xi $ are time and
spatial correlation lengths, respectively. However, the time evolution of
the order parameter is, in general, a non-Markovian process and $\theta _{g}$
turns out to be a new independent critical exponent describing the evolving
of the stochastic dynamic process toward the equilibrium.

In order to evaluate the persistence probability $P(t)$, we first define $%
\rho (t)$ as the fraction of runs for which the magnetization changes its
sign for the first time at the instant $t$. Our probability $P(t)$ is
numerically calculated from the cumulative distribution function of such $%
\rho (t)$. So, $P(t)$ describes the probability of magnetization does not
cross the origin up to time $t$, 
\begin{equation}
P(t)=1-\sum\limits_{t^{\prime }=1}^{t}\rho (t^{\prime })\,.
\end{equation}%
We start our simulations with a random initial condition, where $%
\left\langle m_{0}\right\rangle =0$. Here it is important to mention that
such a concept is very versatile and it has been applied to characterize
several applications such as tricritical points \cite{daSilva2003}, besides
interdisciplinary applications such as: analysis of bankruptcies of players
in public goods games \cite{daSilva2006}, Econophysics \cite{daSilva2010}
and many others. So it can also be interesting, for example, to study
Lifshitz points in spin models.

The target of this paper is to enlarge our knowledge of the of the ANNNI
model, by studying the ferromagnetic-paramagnetic phase-transition with
special attention to the Lifshitz point (LP). The layout of this paper is as
follows: In section (\ref{MCsimulations}) we give more details about
numerical simulations that will be performed at the Lifshitz point.
Moreover, we describe a simple method recently developed by R. da Silva et
al. in \cite{daSilva2012}\ that refines the critical parameters based on the
best determination coefficient in linear fit $\ln \langle M\rangle $ versus $%
\ln t$. In this same section we present a first part of our results, where
we explicitly show the refinement of critical temperatures of second order
line (ferromagnetic-paramagnetic) up to the Lifshitz point. Finally in
section (\ref{Results}) we present our estimates for the critical exponents
and in this case we divide our results in two branches: In the branch (A),
we evaluate critical exponents (only $z$ and $\beta /\nu $) for each
temperature estimated along critical line previously estimated by the
refinement process. In these first calculations, the aim is only the
monitoring of the behavior of these two exponents (one dynamic and the other
static) to show the pronounced difference between Ising-like points and the
Lifshitz point. In the second branch (B) of this same section we present the
complete results and studies for both dynamic ($\theta $, $\theta _{g}$, and 
$z$) and static ($\beta $ and $\nu $) critical exponents specifically for
the Lifshitz point, comparing the later with results obtained in previous
experimental and theoretical works. Finally, in section (\ref{Conclusions})
we summarize and briefly discuss the main results of this paper.

\section{Monte Carlo simulations}

\label{MCsimulations}

Monte Carlo simulations were performed on simple rectangular lattices with
linear dimensions $L_{x}$, $L_{y}$, and $L_{z}$ and periodic boundary
conditions. The spin states were updated by using the one-spin-flip
heat-bath algorithm. For the location of the LP, it was used as basis the
result obtained by Henkel and Pleimling \cite{Henkel2002}: ($%
-J_{2}/J_{1}=0.270\pm 0.004;k_{B}T/J_{1}=3.7475\pm 0.005$).

For the ANNNI model the order parameter corresponds to time-dependent
magnetization, defined as an average over all $L_{x}\times L_{y}\times L_{z}$
spins and over the different $N_{s}$ runs:

\begin{equation}
\left\langle M\right\rangle (t)=\frac{1}{n_{s}L^{3}}\sum%
\limits_{i=1}^{n_{s}}\sum\limits_{x,y,z}\sigma _{x,y,z}^{(i)}(t)
\label{magnetizacao}
\end{equation}%
where the index $i=1,...,N_{s}$ denotes the corresponding run of a
simulation. The ordered state is ferromagnetic, with all (or most of) the
spins pointing either up or down.

As discussed in the previous section, the lattice's initial condition
depends on the scaling relation considered: a) $\left\langle
m_{0}\right\rangle =0$ -- Eq. (\ref{correlation}), b) $m_{0}=0$ -- Numerator
of Eq. (\ref{z}), c) $m_{0}=1$ Eqs.(\ref{decay_ferro}), (\ref{aproximation}%
), and d) $m_{0}$ fixed, but with random configurations considering a sharp
preparation (\ref{initial slip}).

Here we address time dependent MC simulations in the context of so-called
short time dynamics. Before evaluation of some critical exponents and the
complete study of the Lifshitz point (next section), also in this section we
apply the refinement method to estimate the critical parameters of several
points along second order line including the Lifshitz point itself.

The algorithm to estimate the critical temperature is divided in two stages:
i) a coarse grained location; ii) fine scale refinement. In stage i), since
the magnetization must behave as a power law $\langle M\rangle \sim
t^{-\beta /\nu z}$, by fixing a specific $\alpha $-value, we conjecture that
changing $J_{1}/k_{B}T_{c}$ from $J_{1}/k_{B}T_{c}^{(\max )}$ up to $%
J_{1}/k_{B}T_{c}^{(\min )}$, the corresponding best $%
J_{1}/k_{B}T_{c}^{(best)}$ is the one that leads to the best linear behavior
of $\ln \langle M\rangle $ versus $\ln t$. We have considered $n_{s}=400$
realizations, with initial magnetization $m_{0}=1$. For each $\alpha
=-J_{2}/J_{1}$ changing from $0$ up to $0.27$, with displacement $\Delta
\alpha =0.03$ between the values, where $J_{0}=J_{1}$ for all of our
calculations, we changed $J_{1}/k_{B}T_{c}$ from $J_{1}/k_{B}T_{c}^{(\max
)}=0.26684456...$ up to $J_{1}/k_{B}T_{c}^{(\min )}=0.22165413...$. These
extreme values were extracted from the literature, since they correspond to
best known estimates for the 3D Ising model ($\alpha =0$ which corresponds
to $k_{B}T_{c}/J=4.5115333351...$) and LP ($\alpha =0.27$ which corresponds
to $k_{B}T_{c}/J=3.7475..$) respectively. Just as a safeguard, we enlarge
this interval, by performing $J_{1}/k_{B}T_{c}^{(\max ,0)}=0.28$ and $%
J_{1}/k_{B}T_{c}^{(\min ,0)}=0.21$.

So, for each input $\alpha $-value, by using $\Delta
^{(0)}(J_{1}/k_{B}T_{c})=0.002$, we span our temperatures on a range
described by parametrization $J_{1}/k_{B}T_{c}=J_{1}/k_{B}T_{c}^{(\min )}+j$ 
$\Delta ^{(0)}(J_{1}/k_{B}T_{c})$, $j=0,...,32$, so for each temperature, a
linear fit is performed and one calculates the determination coefficient of
the fit as: 
\begin{equation}
r=\frac{\sum\limits_{t=1}^{N_{MC}}(\overline{\ln \langle M\rangle }-a-b\ln
t)^{2}}{\sum\limits_{t=1}^{N_{MC}}(\overline{\ln \left\langle M\right\rangle 
}-\ln \langle M\rangle (t))^{2}}
\end{equation}%
with $\overline{\ln \langle M\rangle }=(1/N_{MC})\sum%
\nolimits_{t=1}^{N_{MC}}\ln \langle M\rangle (t)$, where $N_{MC}$ is the
number of Monte Carlo sweeps. In our experiments, we have used $N_{MC}=150$
MC steps, discarding the initial $30$ MCsteps for more robust estimates.
Hence, $r=1$ means an exact fit, and so the closer $r$ is from the unity,
the better is the fit. Here, $a$ and $b$ are the linear coefficient and the
slope in the linear fit $\ln \langle M\rangle $ versus $\ln t$,
respectively. From $b$, one estimates the exponent $\beta \nu /z$.

After, one finishes this part of our refinement method (i), we pass to the
second part of refinement, the fine scale stage (ii). Starting from the
critical temperature $k_{B}T_{c}^{(1)}(q)/J_{1}$ obtained in the first
stage, we use the process considering a more refined displacement, i.e., $%
\Delta ^{(1)}(J_{1}/k_{B}T_{c})=1\cdot 10^{-4}$. So, by using $%
J_{1}/k_{B}T_{c}^{(\min ,1)}=J_{1}/k_{B}T_{c}^{(1)}-\Delta
^{(0)}(J_{1}/k_{B}T_{c})$ and $J_{1}/k_{B}T_{c}^{(\max
,1)}=J_{1}/k_{B}T_{c}^{(1)}+\Delta ^{(0)}(J_{1}/k_{B}T_{c})$, we consider a
new parametrization $J_{1}/k_{B}T_{c}=J_{1}/k_{B}T_{c}^{(\min ,1)}+$ $\Delta
^{(1)}(J_{1}/k_{B}T_{c})j$, with $j=0,...,41$. Hence, we determine a new
best temperature $k_{B}T_{c}^{(2)}/J_{1}$ corresponding to the maximum value
of $r$.

\begin{figure}[h]
\begin{center}
\includegraphics[width=\columnwidth]{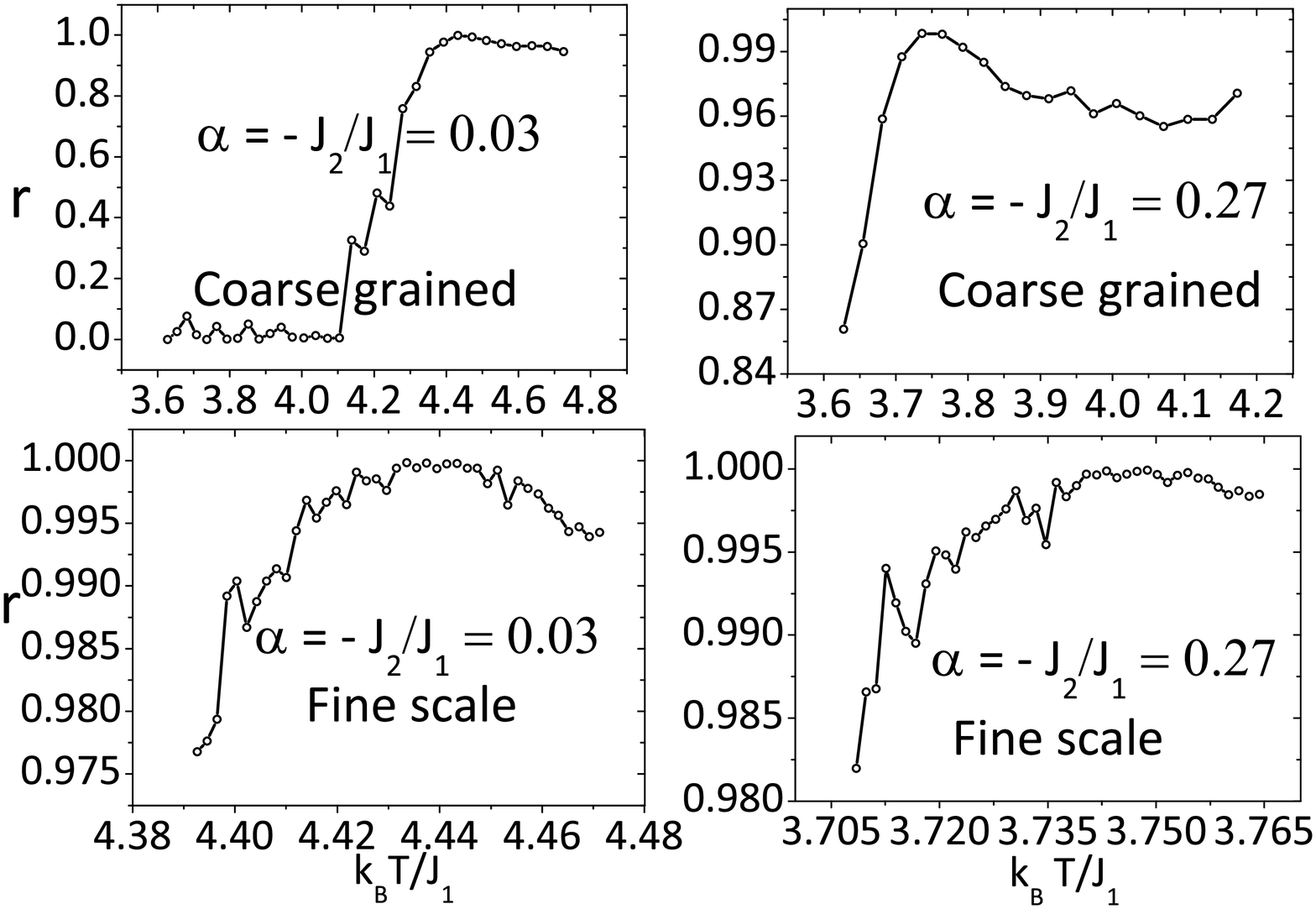}
\end{center}
\caption{Determination coefficient ($r$) as function of temperature for two
cases $\protect\alpha =0.03$ and $\protect\alpha =0.27$. The upper plots
correspond to coarse grained regime (i) of process ($\Delta
(J_{1}/k_{B}T)=2.10^{-3}$) while the lower plots correspond to fine scale
regime (ii) of process ($\Delta (J_{1}/k_{B}T)=1.10^{-4}$). }
\label{refinement}
\end{figure}

Plot \ref{refinement} shows the determination coefficient ($r$) as function
of temperature for two extremal cases $\alpha =0.03$ and $\alpha =0.27$.

The upper plots correspond to the coarse grained regime (first refinement --
i) of process ($\Delta (J_{1}/k_{B}T)=2.10^{-3}$) while the lower plots
correspond to the fine scale regime (second refinement -- ii) ($\Delta
(J_{1}/k_{B}T)=1.10^{-4}$). The estimated temperatures after two refinement
states for all studied $\alpha $-values can be observed in table \ref%
{table:critical_line_annni}.

\begin{table}[tbp]
\centering
\begin{tabular}{ccc}
\hline\hline
$\alpha =-J_{2}/J_{1}$ & $k_{B}T_{c}/J_{1}$ & $r$ \\ \hline
\textbf{0.00} & \textbf{4.513} & \textbf{0.99993} \\ 
0.03 & 4.434 & 0.99982 \\ 
0.06 & 4.364 & 0.99990 \\ 
0.09 & 4.289 & 0.99990 \\ 
0.12 & 4.213 & 0.99993 \\ 
0.15 & 4.126 & 0.99991 \\ 
0.18 & 4.039 & 0.99988 \\ 
0.21 & 3.956 & 0.99991 \\ 
0.24 & 3.852 & 0.99989 \\ 
\textbf{0.27} & \textbf{3.748} & \textbf{0.99992} \\ \hline\hline
\end{tabular}%
\caption{Critical values obtained by refinement procedure after two stages
(coarse-grained and fine scale). Here, the values obtained for $\protect%
\alpha =0$ and $\protect\alpha =0.27$ correspond to our estimates of the 
critical temperatures of 3D-Ising model and Lifshtz point respectively.}
\label{table:critical_line_annni}
\end{table}

Since our final refinement has a precision of $\Delta
(J_{1}/k_{B}T_{c})=\Delta (\beta _{c})=10^{-4}$, which means $\Delta
(T_{c})=\Delta (\beta _{c})T_{c}^{2}/(1+\Delta (\beta _{c})T_{c})$, we have
a precision with 3 digits for temperature. Therefore, we show our results
for critical temperatures with 3 significant elements (second column in
table \ref{table:critical_line_annni}). It is important to mention that our
estimates corroborate literature results: For example, for $\alpha =0$ we
have as 3D Ising estimate $k_{B}T_{c}/J_{1}=4.513$ which yields an excellent
agreement with estimates via equilibrium Monte Carlo simulations \cite%
{Janke1997}. For $\alpha =0.27$, we have $k_{B}T_{c}/J_{1}=3.748$ as best
estimate for the LP's critical temperature, which also agrees with the
estimate obtained by Henkel and Pleimling \cite{Henkel2002} $%
k_{B}T/J_{1}=3.7475\pm 0.0050$.

\begin{figure}[h]
\begin{center}
\includegraphics[width=\columnwidth]{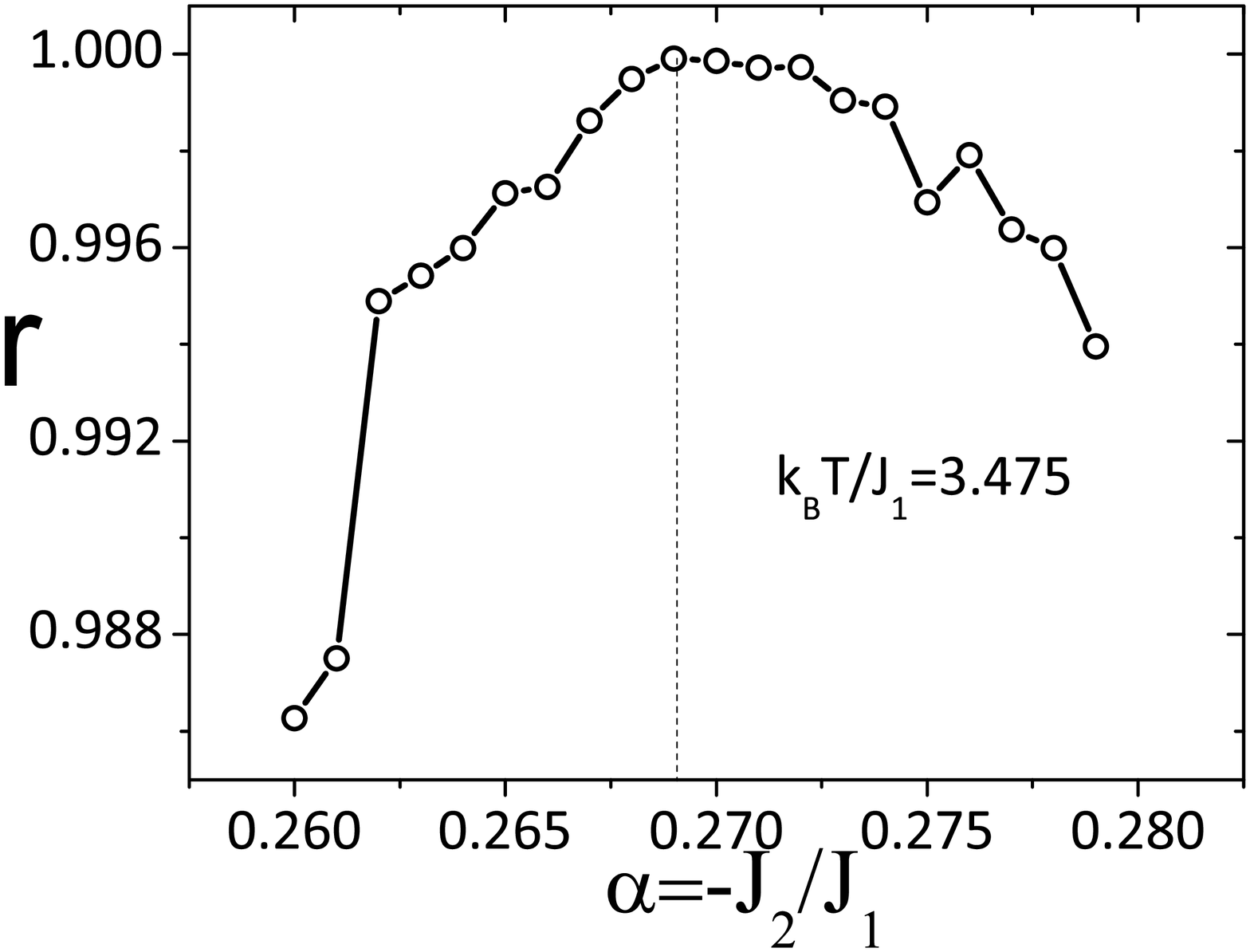}
\end{center}
\caption{\textbf{Inverted refinement process}: determination coefficient ($r$%
) as function of $\protect\alpha $, which changes from $0.26$ up to $0.28$
by fixing $k_{B}T/J_{1}=3.475$. Only one refinement process with $\Delta 
\protect\alpha =0.001$ was used. We obtain as best estimate $\protect\alpha %
=0.269$, which shows that the method works in both ways. }
\label{refinement2}
\end{figure}

Here, it is important to mention that in order to check the robustness of
our method, we performed the same refinement process inverting the input,
i.e., we fixed $k_{B}T/J_{1}=3.7475$ and we refine the value $\alpha $. For
the sake of simplicity, here we perform only one refinement in a shorter
interval spanning $\alpha -$values from $0.26$ to $0.28$ with $\Delta \alpha
=0.001$ which can be observed in Fig. \ref{refinement2}.

We find $\alpha =0.269$ as the best estimate which corroborates the expected
value for $k_{B}T/J_{1}=3.7475$: $\alpha =0.270(4)$.

So, once we have shown that our time-dependent simulations are calibrated
and they corroborate the critical parameters of the main estimates of
literature, in the next section, our focus is to calculate the critical
exponents via time-dependent Monte Carlo simulations of LP. First, we
present some exponents ($z$ and $\beta /\nu $) to monitor their behavior
along the second order transition. After, we show a detailed study for the
Lifshitz point, by calculating all (static an dynamic) critical exponents
obtaining error bars by performing repetitions of simulations under
different seeds. We also studied some differences between rectangular and
cubic lattices.

\section{Results}

\label{Results}

In this work we perform short-time Monte Carlo simulations in simple
rectangular lattices of size $L_{x}\times L_{y}\times L_{z}$ and not only
for $L_{x}=L_{y}=L_{z}=L$. Thus, at each instant $t$ of the simulation, the
value of any measured quantity is given by its average over $n_{s}$ runs
according to equation (\ref{magnetizacao}), which denotes an average over
different repetitions (trajectories with different sequences of
pseudo-random numbers).

\begin{figure}[h]
\begin{center}
\includegraphics[width=\columnwidth]{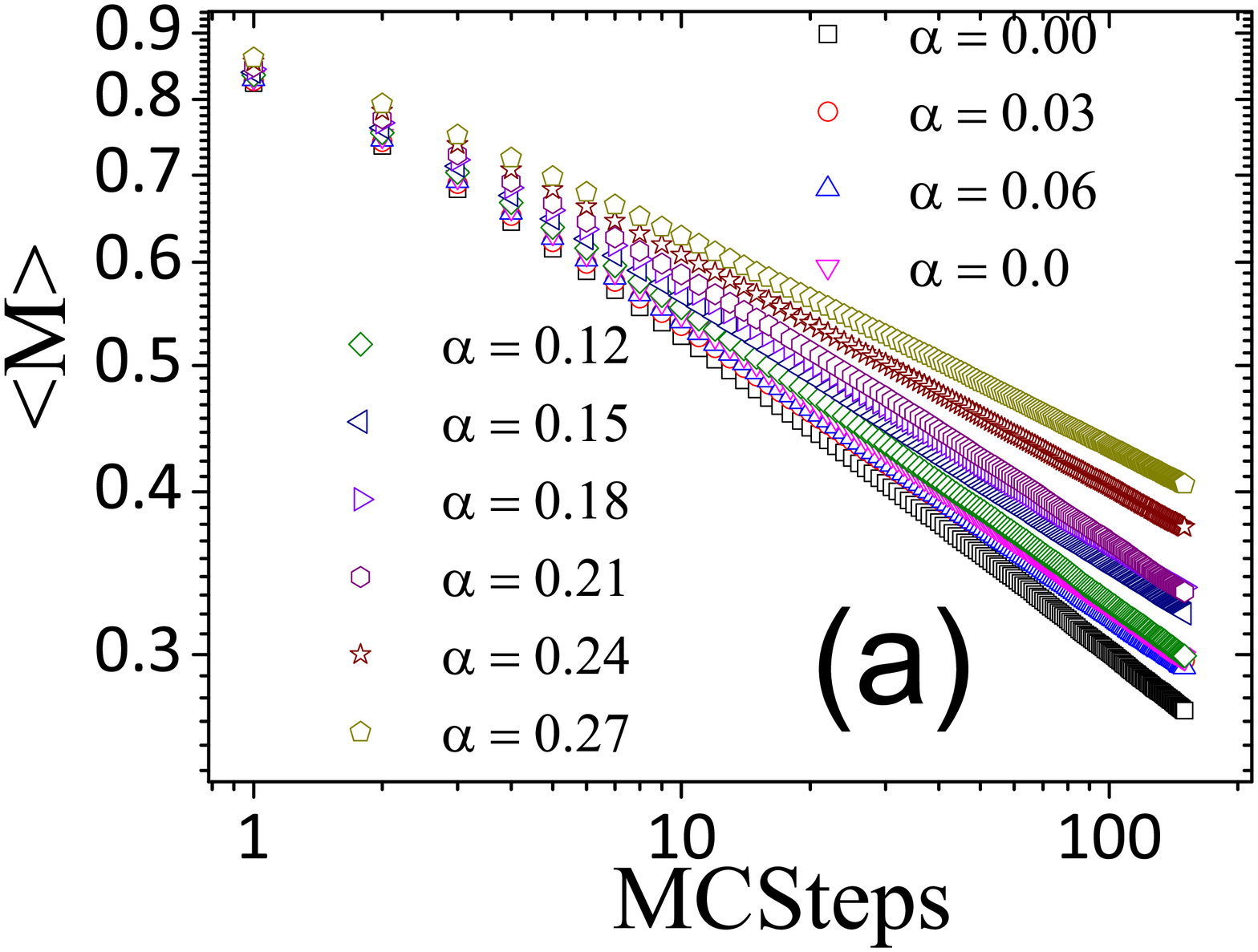} \includegraphics[width=%
\columnwidth]{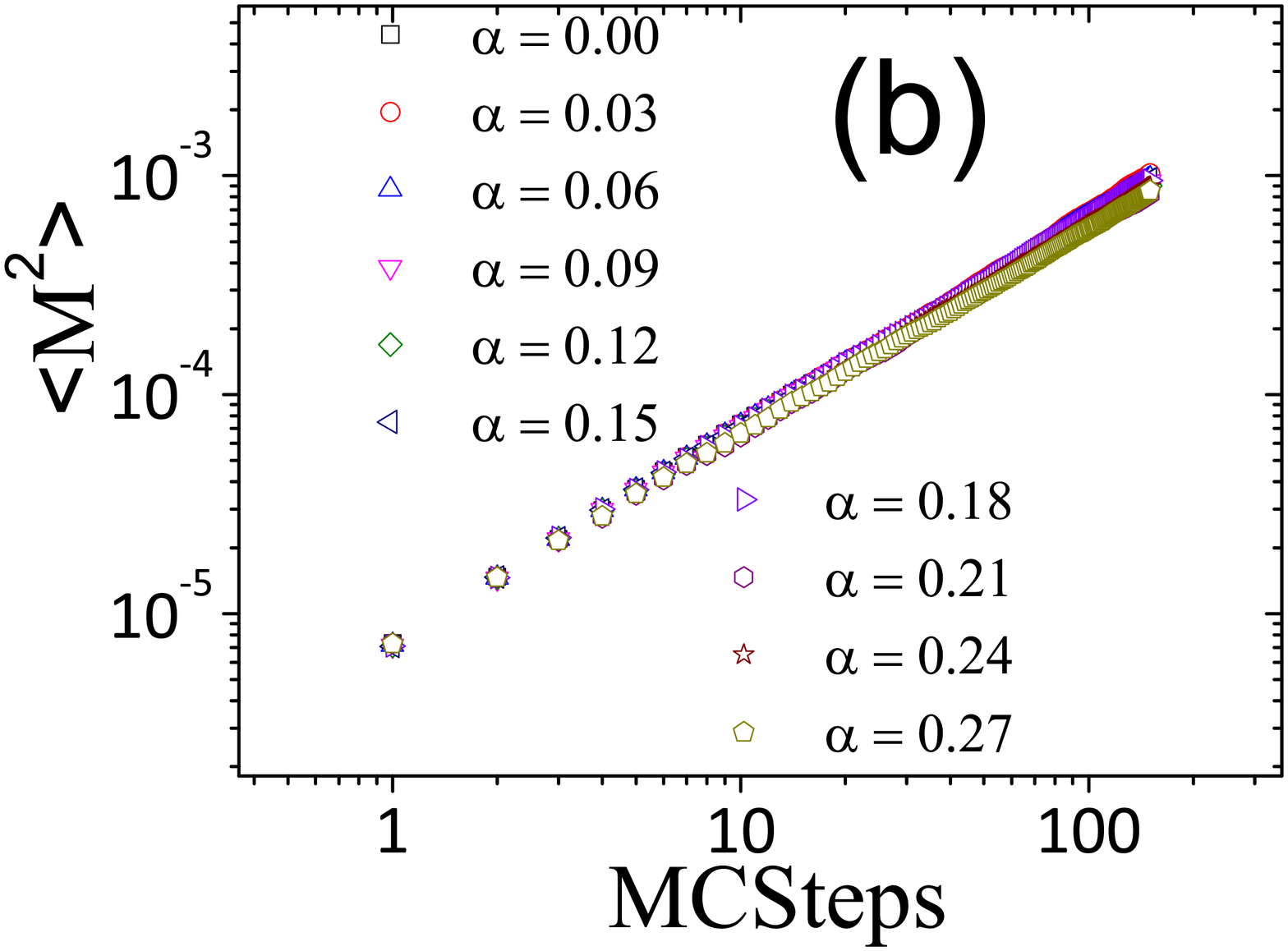} \includegraphics[width=\columnwidth]{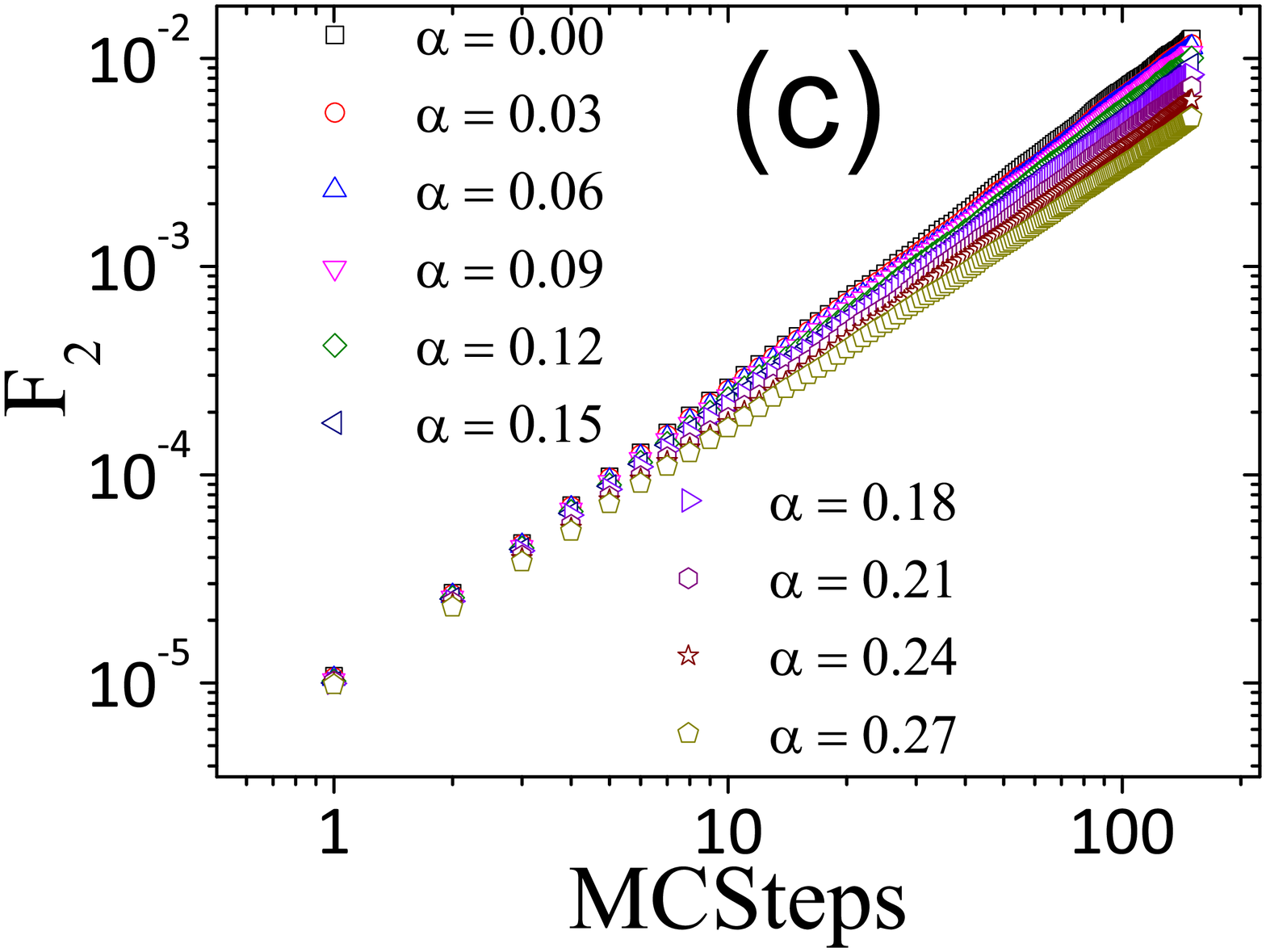}
\end{center}
\caption{(Color Online) Time evolving of $\left\langle M(t)\right\rangle
_{m_{0}=1}$(plot a), $\left\langle M^{2}(t)\right\rangle _{m_{0}=0}$(plot
b), and $F_{2}(t)$ (plot c) for all critical temperatures previously
estimated (corresponding to the different $\protect\alpha $-values) }
\label{time_evolving}
\end{figure}

In order to obtain error bars in our simulations for the LP, we used $%
n_{b}=5 $ sets of $n_{s}\ $runs corresponding to different seeds for the
random numbers generator. For our calculations, we first used cubic lattices
($80\times 80\times 80$) to evaluate only two exponents ($z$ and $\beta /\nu 
$) along the second order line by performing MC simulations using as input
the critical parameters obtained in the previous section. We monitor the
behavior of these two exponents directly obtained as a function of $%
k_{B}T_{c}/J_{1}$.

Second, for the Lifshitz point we used three different lattices $80\times
80\times 80$, $80\times 80\times 10$, and $120\times 120\times 10$ in order
to study the possible distortions under rectangular regions. For each
lattice size, the number of runs used in the simulations in order to obtain
the exponents given in Eqs. (\ref{decay_ferro}), (\ref{M2}), (\ref%
{derivative}), (\ref{correlation}), (\ref{initial slip}), and (\ref%
{global_persistence}) are shown in table \ref{tables:runs}.

\begin{table}[tbp]
\centering%
\begin{tabular}{ccccccc}
\hline\hline
&  & $80\times 80\times 80$ &  & $80\times 80\times 10$ &  & $120\times
120\times 10$ \\ \hline\hline
\multicolumn{1}{l}{$\beta /\nu z$. Eq.(\ref{decay_ferro})} & 
\multicolumn{1}{l}{} & $400$ &  & $3200$ &  & $1400$ \\ 
\multicolumn{1}{l}{$(d-\frac{2\beta }{\nu })/z$. Eq.(\ref{M2})} & 
\multicolumn{1}{l}{} & $4000$ &  & $3.2\cdot 10^{4}$ &  & $1.4\cdot 10^{4}$
\\ 
\multicolumn{1}{l}{$1/\nu z$. Eq.(\ref{derivative})} & \multicolumn{1}{l}{}
& $2000$ &  & $1.6\cdot 10^{4}$ &  & $7000$ \\ 
\multicolumn{1}{l}{$\theta $, Eq.(\ref{initial slip})} & \multicolumn{1}{l}{}
& $14000$ &  & $14000$ &  & $14000$ \\ 
\multicolumn{1}{l}{$\theta $, Eq.(\ref{correlation})} & \multicolumn{1}{l}{}
& $28000$ &  & $28000$ &  & $28000$ \\ 
\multicolumn{1}{l}{$\theta _{g}$, Eq.(\ref{global_persistence})} & 
\multicolumn{1}{l}{} & $14000$ &  & $14000$ &  & $14000$ \\ \hline\hline
\end{tabular}%
\caption{Number of runs used in simulations for different size lattices}
\label{tables:runs}
\end{table}

Due to the axial anisotropy present in the ANNNI model, the critical
behavior in the neighborhood of the LP is governed by two correlation
lengths, $\xi _{z}$ and $\xi _{xy}$ with different critical exponents ($\nu
_{z}=\frac{1}{2}\nu _{xy}$) \cite{Horneich1975}. Therefore, at the LP, in
the case of the simple cubic lattice, the resulting critical exponent $\nu $
may be either the exponent $\nu _{z}(\nu _{xy})$ or a combination of these
exponents. Therefore, in order to calculate the static critical exponent $%
\nu $ for the Lifshitz point, we also considered not only cubic lattices in
our study but also rectangular lattices.

\subsection{Monitoring the critical exponents of the ANNNI model}

Using as input the critical parameters previously estimated in section (\ref%
{MCsimulations}), we have calculated the exponent $z$ from the
time-dependence of the ratio $F_{2}$ (Eq.(\ref{z})) for cubic lattices with $%
L=80$. Since $z$ was calculated, the time evolving from ordered state (Eq. %
\ref{decay_ferro}) which was used to compose $F_{2}$ is taken also to obtain
an estimate of ($\beta /\nu z$) and since we have an estimate of $z$ by
equation Eq.(\ref{eta}) we obtain an estimate of $\beta /\nu $.

Performing MC simulations up to 150 MCsteps and discarding the initial 30
MCsteps for more robust estimates we estimate these exponents for all $%
\alpha -$values corresponding to the critical temperatures which were
previously obtained in section (\ref{MCsimulations}), considering the
limits: 3D-Ising like ($\alpha =0$) up to the Lifshitz point ($\alpha =0.27$%
). We set our simulations exactly at the temperatures obtained by our
refinement procedure. The time evolving of the magnetization $\left\langle
M(t)\right\rangle _{m_{0}=1}$, of its second moment $\left\langle
M^{2}(t)\right\rangle _{m_{0}=0}$ and of $F_{2}(t)$ can be observed in
figure \ref{time_evolving}.

The estimates of $z$ and $\beta /\nu $ are summarized in table \ref%
{Table:Monitoring}. We can observe that $z$ is universal when $J_{2}$ is
small, i.e., the interaction among second neighbors in $z$ direction is not
pronounced and $z\approx 2.06$ which is expected for universality class of
the 3D Ising model.

\begin{table}[tbp] \centering%

\begin{tabular}{ccccc}
\hline\hline
$\alpha =-J_{2}/J_{1}$ &  & $z$ &  & $\beta /\nu $ \\ \hline\hline
\multicolumn{1}{l}{\textbf{0.00}} &  & \multicolumn{1}{l}{\textbf{2.068}} & 
& \multicolumn{1}{l}{\textbf{0.5118}} \\ 
\multicolumn{1}{l}{0.03} &  & \multicolumn{1}{l}{2.069} &  & 
\multicolumn{1}{l}{0.4465} \\ 
\multicolumn{1}{l}{0.06} &  & \multicolumn{1}{l}{2.061} &  & 
\multicolumn{1}{l}{0.4807} \\ 
\multicolumn{1}{l}{0.09} &  & \multicolumn{1}{l}{2.076} &  & 
\multicolumn{1}{l}{0.4777} \\ 
\multicolumn{1}{l}{0.12} &  & \multicolumn{1}{l}{2.086} &  & 
\multicolumn{1}{l}{0.4746} \\ 
\multicolumn{1}{l}{0.15} &  & \multicolumn{1}{l}{2.137} &  & 
\multicolumn{1}{l}{0.4478} \\ 
\multicolumn{1}{l}{0.18} &  & \multicolumn{1}{l}{2.172} &  & 
\multicolumn{1}{l}{0.4334} \\ 
\multicolumn{1}{l}{0.21} &  & \multicolumn{1}{l}{2.197} &  & 
\multicolumn{1}{l}{0.4676} \\ 
\multicolumn{1}{l}{0.24} &  & \multicolumn{1}{l}{2.290} &  & 
\multicolumn{1}{l}{0.4082} \\ 
\multicolumn{1}{l}{\textbf{0.27}} &  & \multicolumn{1}{l}{\textbf{2.338}} & 
& \multicolumn{1}{l}{\textbf{0.3867}} \\ \hline\hline
\end{tabular}%
\caption{Monitoring of critical exponent $z$ and the ratio of exponents $\beta/\nu$ along the second order line
previously estimated by time-dependent MC simulations}\label%
{Table:Monitoring}%
\end{table}%

However, in the neighborhood of the Lifshitz point, the exponent $z$ has a
sensitive increasing in relation to Ising like behavior. The ratio of
exponents $\beta /\nu $ changes in an interval [$0.38,0.52$] but not
monotonically as occurs with $z$. In the next subsection we will study all
details of the Lifshitz point, taking into consideration error bars obtained
by considering simulations with different seeds, since we observed a
notorious difference between this point ($\alpha =0.27$) and the authentic
3D Ising model($\alpha =0.0$).

\subsection{Lifshitz Point}

In this subsection, we finally present a detailed study of critical
exponents of the Lifshitz point. Initially we considered cubic lattices $%
L_{x}=L_{y}=L_{z}=80$. Here it is important to stress that we used $\alpha
=0.27$ and $k_{B}T_{c}/J_{1}=3.7475$ (estimate obtained by Henkel and
Pleimling \cite{Henkel2002}) to perform the simulations. 

\begin{figure}[h]
\begin{center}
\includegraphics[width=1.2\columnwidth]{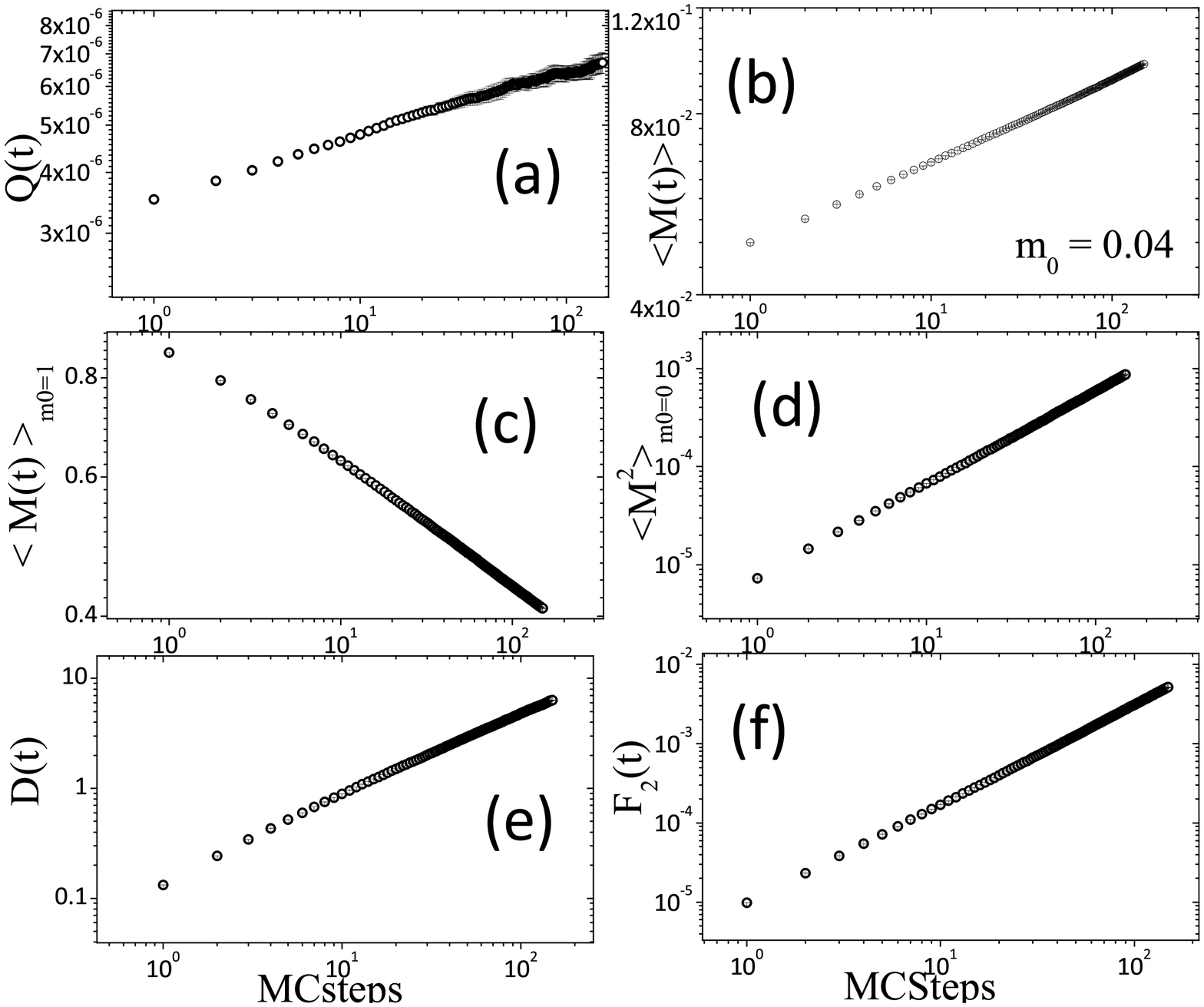}
\end{center}
\caption{(a) log-log curve of $Q(t)$ against $t$, for the lattice size $%
80\times 80\times 80$ for the Lifshitz point by using $\protect\alpha =0.27$
and $k_{B}T_{c}/J_{1}=3.7475$. The plots (b), (c), (d), (e) and (f)
correspond to same plots plot for $\left\langle M(t)\right\rangle
_{m_{0}=0.04}$, $\left\langle M(t)\right\rangle _{m_{0}=1}$, $\left\langle
M^{2}(t)\right\rangle _{m_{0}=0}$,$\ D(t)$, and $F_{2}(t)$.}
\label{figure_all_plots}
\end{figure}

So, first of all, from Monte Carlo simulations we obtained the exponents for
the Lishitz point defined by equations (\ref{z}) (that uses the equations (%
\ref{decay_ferro}) and (\ref{M2})), (\ref{derivative}), (\ref{initial slip}%
), and (\ref{correlation}). The log-log curves for the Lifshitz point for
these quantities considering the error bars obtained by 5 different seeds
are shown in Fig.(\ref{figure_all_plots}). For equation (\ref{initial slip})
we show (for the sake of simplicity) only the evolution for $m_{0}=0.04$.

\begin{figure}[h]
\begin{center}
\includegraphics[width=\columnwidth]{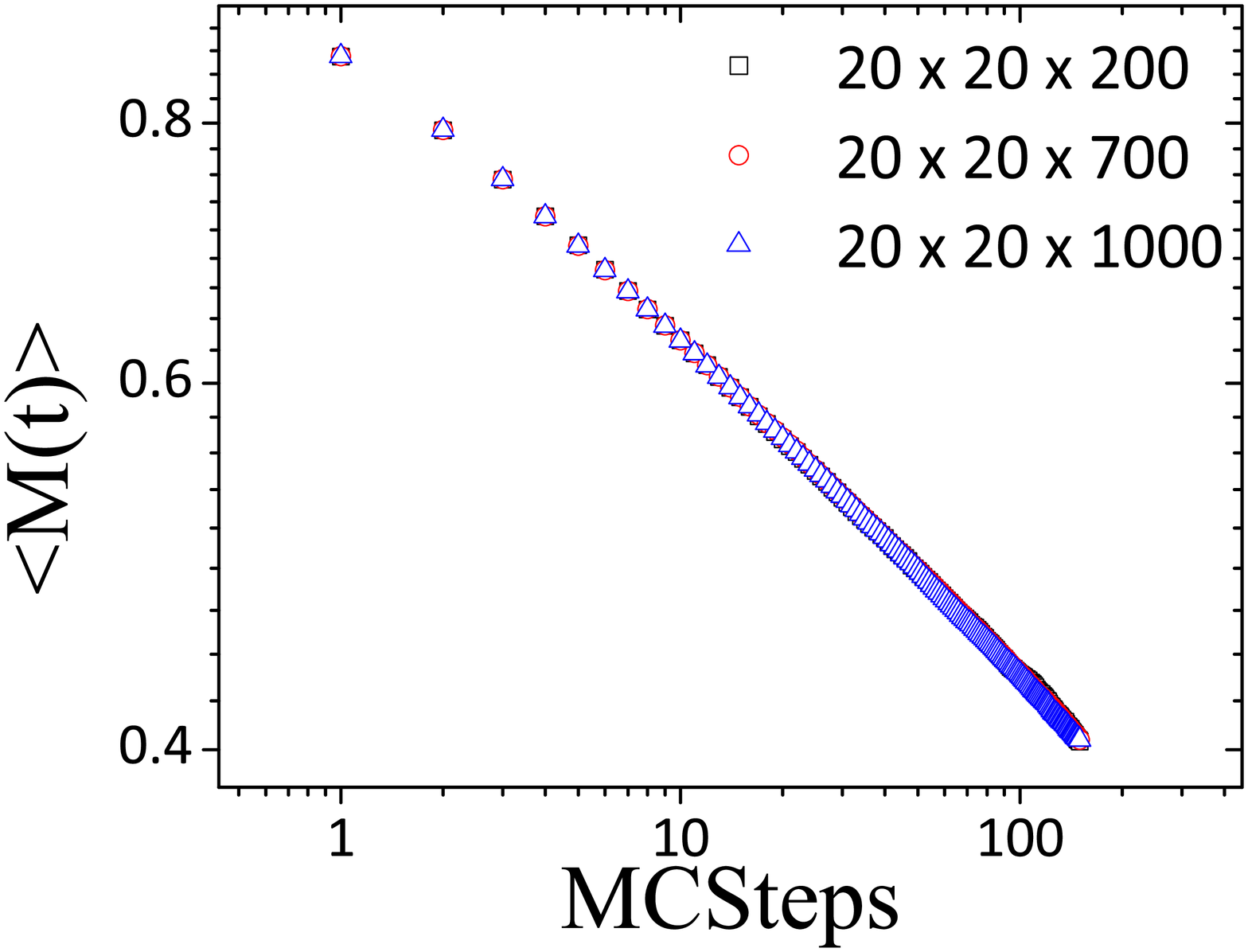}
\end{center}
\caption{(Color Online) Log-log curves of $\left\langle M(t)\right\rangle $
against $t$ obtained from the simulations performed for the lattice sizes $%
20\times 20\times L_{z}$, where $L_{z}>20$ and $m_{0}=1$(ordered initial
state). }
\label{retangular}
\end{figure}

We have obtained the critical exponents $\beta =0.226(2)$, $\nu =0.60(1)$, $%
z=2.34(2)$, $\theta =0.17(2)$ (from Eq. \ref{correlation}) and $\theta
=0.163(3)$ (Eq. \ref{initial slip}). For this last case the results were
obtained by performing simulations with 3 different initial magnetizations $%
m_{0}=0.02$, $m_{0}=0.04$ and, $m_{0}=0.08$. By obtaining the exponent for
each value with error bars obtained by 5 seeds, we considered a linear fit
to obtain an extrapolation $m_{0}\rightarrow 0$, which corresponds to linear
coefficient in the fit $\theta $ versus $m_{0}$.

To verify whether the ferromagnetic ordering is affected by increasing the
linear dimension $L_{z}$, we perform extra MC simulations for the temporal
evolution of magnetization from ordered initial state $m_{0}=1$, considering
the following rectangular lattices: $20\times 20\times 200$, $20\times
20\times 700$, and $20\times 20\times 1000$. Here we are only interested in
the qualitative plots of $\left\langle M(t)\right\rangle $ versus $t$. We
used in all cases just $n_{s}=400\ $runs. In Fig. \ref{retangular} we find
identical log-log plots for different rectangular lattices analyzed, which
corroborates that ferromagnetic ordering does not depend on linear dimension 
$L_{z}$, despite the increasing of the total number of spins on the lattice.

\begin{figure}[h]
\begin{center}
\includegraphics[width=\columnwidth]{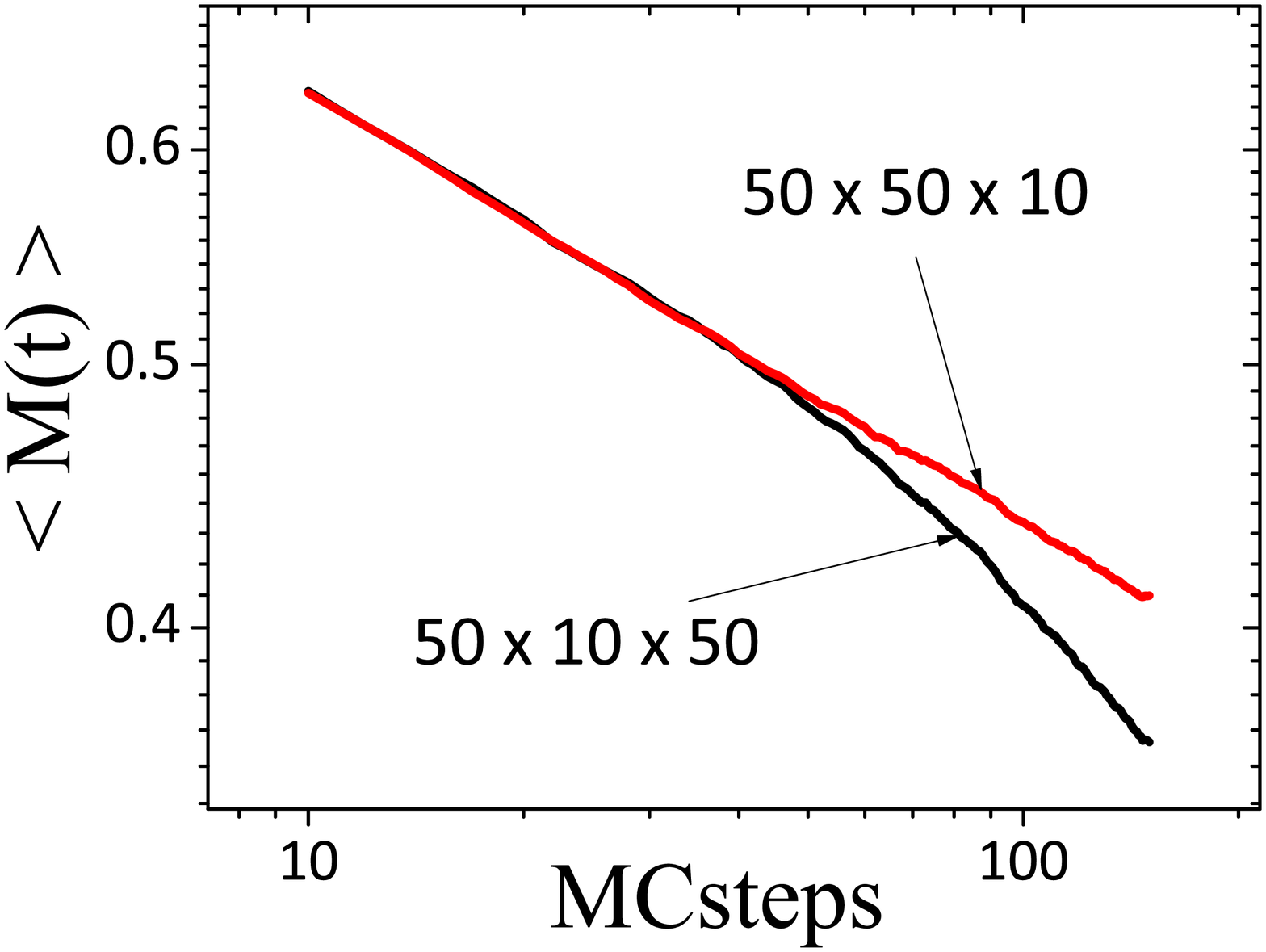}
\end{center}
\caption{(Color online) Log-log curves of $\left\langle M(t)\right\rangle $
against $t$ obtained from the simulations performed for the lattice sizes $%
50\times 50\times 10$ and $50\times 10\times 50$, starting from the initial
condition $m_{0}=1$(ordered initial state). The order parameter decays
faster for the lattice with smaller $xy$ planes. }
\label{xy_effects}
\end{figure}

Since the order parameter is the magnetization and the $xy$ interactions are
ferromagnetic, the ordering occurs mainly due to these interactions. On the
other hand, there are competing interactions along the $z$ axis, which
cannot sustain the ferromagnetic ordering of the system. This fact is
illustrated in Fig. \ref{xy_effects}, where the magnetizations $\left\langle
M(t)\right\rangle $ obtained from simulations for lattices $50\times
50\times 10$ and $50\times 10\times 50$ are shown. In Fig \ref{xy_effects}
we clearly see that the order parameter decays faster for the lattice with a
smaller $xy$ plane ($50\times 10\times 50$).

Therefore, we conclude that the best results will be obtained from lattices
with bigger $xy$ planes, irrespective of the value $L_{z}$. However, we must
stress here, these conclusions are valid in the short-time regime. On the
other hand, via equilibrium MC simulations, which were performed in a simple
cubic lattice ($24\times 24\times 24$), Kaski and Selke \cite{Kaski1985}
obtained the critical exponent $\nu =0.51(4)$ corresponding to a combination
of the exponent $\nu _{z}$ and $\nu _{xy}$ due to anisotropy present in the
LP. In order to obtain the critical exponent $\nu _{z}=0.33(3)$ shown in
table \ref{static_critical_exponents} that show exponents for our comparison
extracted from literature obtained for different methods, these authors
divided the lattice into sub-cells $24\times 24\times L$. By using this
procedure, Kaski and Selke \cite{Kaski1985} were able to capture the spacial
correlations $\xi _{z}$ and from the slope of Binder's cumulant \cite%
{Binder1981,BinderII1981}, they estimated the value of $\nu _{z}$ which is
also shown in table \ref{static_critical_exponents}.

\begin{table}[tbp] \centering%
\begin{tabular}{lllll}
\hline\hline
& 
\begin{tabular}{l}
$1\text{st}\ \text{order}\ \varepsilon $ \\ 
$\text{\cite{Horneich1975}}$%
\end{tabular}
& $%
\begin{tabular}{l}
2nd order $\varepsilon $ \\ 
\cite{Shpot2001}%
\end{tabular}%
$ & Monte Carlo & 
\begin{tabular}{l}
MnP \\ 
\cite{Shapira1984}%
\end{tabular}
\\ \hline\hline
$\beta $ & 1/4 & 0.220 & $%
\begin{tabular}{l}
0.19(2) \cite{Kaski1985} \\ 
0.238(5) \cite{Henkel2002}%
\end{tabular}%
$ &  \\ 
$\nu _{z}$ & 5/16 & 0.348 & 0.33(3) & 0.30(2) \\ 
$\nu _{xy}$ & 5/8 & 0.696 & 0.66 \cite{Kaski1985} & 0.60(4) \\ \hline\hline
\end{tabular}%
\newline
\caption{Static critical exponents $\beta$, $\nu_{z}$, and
$\nu_{xy}$=2$\nu_{z}$ at the LP extracted from literature.}\label%
{static_critical_exponents}%
\end{table}%

In the following we show the results obtained in this work for the lattice
sizes $80\times 80\times 10$ and $120\times 120\times 10$. We estimated the
critical exponents $\beta =0.227(1)$, $z=2.296(3)$ and $\nu =0.615(3)$ ($%
80\times 80\times 10$) and $\beta =0.229(2)$, $z=2.30(1)$ and $\nu =0.618(4)$
($120\times 120\times 10$).

\begin{figure}[h]
\begin{center}
\includegraphics[width=\columnwidth]{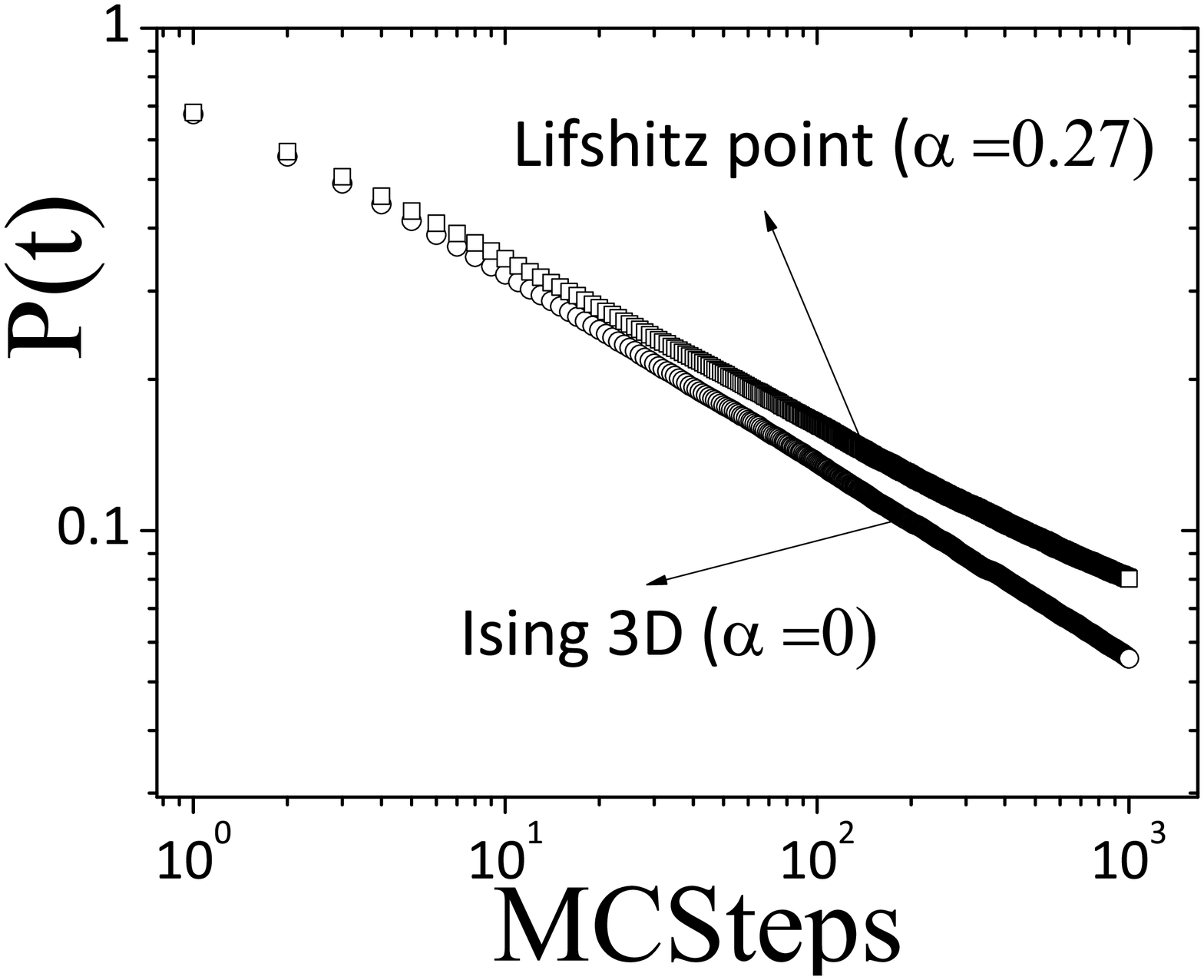}
\end{center}
\caption{Time evolving of global persistence for $\protect\alpha =0$ (%
\textbf{3D\ Ising model}) $k_{B}T/J_{1}=4.513$ and $\protect\alpha =0.27$ (%
\textbf{Lifshitz point}) $k_{B}T/J_{1}=3.7475$. }
\label{persistence_plot}
\end{figure}

Finally, we also study the global persistence for the Lifshitz point. For a
simple comparison, we also performed simulations for $\alpha =0$ (that
corresponds to the three-dimensional Ising model). The same procedure used
in the previous simulations (5 seeds to obtain error bars) was replicated
here as well. The figure \ref{persistence_plot} shows the different time
evolutions of global persistence $P(t)$ for the two different points.

We obtain for lattices $80\times 80\times 80$, $\theta _{g}=0.336(4)$ for
the Lifshitz point. This value is smaller than $\theta _{g}=0.384(6)$ found
for the 3D Ising model. This value for the Lifshitz point seems to be
corroborated for rectangular lattices (compatible with error bars) where we
observed only small changes for rectangular lattices $80\times 80\times 10$
and $120\times 120\times 10$ (see last row in table \ref{our_results}). \ 

Table \ref{our_results} summarizes the results of the static and dynamic
critical exponents obtained in this work. These results are in very good
agreement with the critical exponents obtained in previous works, shown in
table \ref{static_critical_exponents}.

\begin{table}[tbp] \centering%
\begin{tabular}{llllllll}
\hline\hline
& $80\times 80\times 80$ &  &  & $80\times 80\times 10$ &  &  & $120\times
120\times 10$ \\ \hline\hline
$\beta $ & $0.226(2)$ &  &  & $0.227(1)$ &  &  & $0.229(2)$ \\ 
$z$ & $2.34(2)$ &  &  & $2.296(2)$ &  &  & $2.30(1)$ \\ 
$\nu _{xy}$ & $0.60(1)$ &  &  & $0.615(3)$ &  &  & $0.618(4)$ \\ 
$\theta $ & 
\begin{tabular}{l}
$0.17(2)$ Eq. (\ref{correlation}) \\ 
$0.163(3)$ Eq. (\ref{initial slip})%
\end{tabular}
&  &  & 
\begin{tabular}{l}
$0.16(2)$ Eq. (\ref{correlation}) \\ 
$0.180(6)$ Eq. (\ref{initial slip})%
\end{tabular}
&  &  & 
\begin{tabular}{l}
$0.16(2)$ Eq. (\ref{correlation}) \\ 
$0.184(1)$ Eq. (\ref{initial slip})%
\end{tabular}
\\ 
$\theta _{g}$ & $0.336(4)$ &  &  & $0.330(5)$ &  &  & $0.336(6)$ \\ 
\hline\hline
\end{tabular}%
\caption{Static and dynamic critical exponents $\beta$, $z$, $\nu_{xy}$, $\theta$, and $\theta_{g}$ obtained in the present 
work for different lattice sizes.}\label{our_results}%
\end{table}%

\section{Summary and Discussion}

\label{Conclusions}

In this work we estimated the dynamic critical exponents ($\theta $, $%
\theta_{g} $, and $z$) at the Lishshitz point of the ANNNI model, by means
of time dependent Monte Carlo simulations. To our knowledge, this is the
first attempt in order to obtain such exponents.

We have also obtained the static and critical exponents $\beta $ and $\nu $
by exploiting scaling relations, valid in the short-time regime (out of
equilibrium), involving the order parameter and its second moment. Our
estimates (see table \ref{our_results}) are in very good agreement with
previous experimental and theoretical works (see table \ref%
{static_critical_exponents})

Moreover we applied a refinement procedure to estimate several parameters
for the critical points along second order transition line from the 3D Ising
point ($-J_{2}/J_{1}=0$) up to Lifshitz point ($-J_{2}/J_{1}=0.27$) and we
follow the behavior of some exponents along this line. We can observe that $%
z $ is universal when $J_{2}$ is small, i.e., the interaction among second
neighbors in $z$ direction is not pronounced resulting in $z\approx 2.06$
which is exactly expected for universality class of 3D Ising. However, in
the neighborhood of the Lifshitz point, the exponent $z$ has a sensitive
increase resulting in $z\approx 2.34$. Finally, we also observe power law
behavior for the global persistence ($P(t)$, probability that the sign of
the magnetization does not change until the time $t$ starting from random
configurations with small magnetization). We find $\theta _{g}=0.336(4)$ ($%
\alpha =0.27$) for the LP that is smaller than 3D-Ising estimate ($\alpha =0$%
) $\theta _{g}=0.384(6)$.

\section*{Acknowledgements}


The authors are partly supported by the Brazilian Research Council CNPq.
Authors also thank CESUP (Super Computer Center of Federal University of Rio
Grande do Sul) as well as Prof. Leonardo G. Brunet (IF-UFRGS) for the
available computational resources and support of Clustered Computing
(ada.if.ufrgs.br) and Prof. Mendeli H. Vainstein (IF-UFRGS) for carefully
reading our manuscript.


\end{document}